\documentclass[prb,preprint,showpacs]{revtex4}
\usepackage{graphicx}
\usepackage{amsmath,bm}
\usepackage{pstricks}

\begin{document}
\title{
Quasiharmonic elastic constants corrected for deviatoric thermal stresses 
}

\author{\firstname{Pierre} \surname{Carrier}} 
\affiliation{Minnesota Supercomputing Institute, Department of Chemical Engineering and 
Materials Science, University of Minnesota, Minneapolis, MN 55455}
\author{\firstname{Jo\~{a}o} \surname{F. Justo} } 
\affiliation{Escola Polit\'ecnica, Universidade de S\~{a}o Paulo, CP 61548, 
CEP 05424-970, S\~{a}o Paulo, Brazil, and 
Chemical Engineering and Materials Science Department, University of Minnesota, MN 55455}
\author{\firstname{Renata} \surname{M. Wentzcovitch}} 
\affiliation{Minnesota Supercomputing Institute, Department of Chemical Engineering and 
Materials Science, University of Minnesota, Minneapolis, MN 55455}

\date{\today}
 
\begin{abstract}
The quasiharmonic approximation (QHA), in its simplest form also called the statically 
constrained (SC) QHA, has been shown to be a straightforward method to compute thermoelastic 
properties of crystals.  Recently we showed that for non-cubic solids SC-QHA calculations develop 
deviatoric thermal stresses at high temperatures. Relaxation of these stresses leads to a series 
of corrections to the free energy that may be taken to any desired order, up to self-consistency.
Here we show how to correct the elastic constants obtained using the SC-QHA. We exemplify the 
procedure by correcting to first order the elastic constants of MgSiO$_3$-perovskite and 
MgSiO$_3$-post-perovskite, the major phases of the Earth's lower mantle. We show that this first 
order correction is quite satisfactory for obtaining the aggregated elastic averages of these 
minerals and their velocities in the lower mantle.  This type of correction is also shown to be 
applicable to experimental measurements of elastic constants in situations where deviatoric 
stresses can develop, such as in diamond anvil cells. 
\end{abstract}

\maketitle

\section{Introduction}

The quasiharmonic approximation (QHA) \cite{Wallace, Anderson} is a computationally efficient 
method for evaluating thermal properties of materials within the density functional theory (DFT)
from low to temperatures above the Debye temperature. It provides high quality 
high pressure-high temperature materials properties \cite{karki99, RenataPRL, Sha, Menendez, Tsuchiya, RenataPNAS} in a 
continuous pressure-temperature (PT) domain in which anharmonic effects are 
negligible.\cite{Carrier}
However, it has a not well recognized shortcoming: the non-hydrostatic nature of thermal 
stresses in non-isotropic structures. Broadly speaking, these calculations start by obtaining 
the static internal energy of fully relaxed DFT structures at various pressures. After 
computations of the vibrational density of states, the thermal energy contribution to the 
Helmholtz free energy is added. This latter contribution has anisotropic strain gradients and 
produces deviatoric stresses. This straightforward procedure should be referred to as the 
statically constrained (SC) QHA. It has been used to compute the elastic constant tensor of 
isotropic \cite{karki99} and non-isotropic minerals \cite{RenataPRL, Tsuchiya} at high PT as 
well, even though pressure conditions were not precisely hydrostatic in the latter calculations. 
In general, relaxation of deviatoric stresses, irrespective of their origin, is essential in 
both experiments\cite{Bassett, Menendez} and theory,\cite{Carrier} for generating realistic 
and reproducible structural and elastic properties. 

Here we show how to correct the elastic constant tensor obtained using the SC-QHA. We exemplify 
the procedure by correcting to first order the elastic constants of MgSiO$_3$-perovskite (PV) and 
MgSiO$_3$-post-perovskite (PPV), the major phases of the Earth's lower mantle, for which 
elasticity \emph{data} are essential to interpret seismic information of 
this region.\cite{Lay} We 
show that this first order correction is quite satisfactory for obtaining the aggregated 
elastic averages of these minerals and their acoustic velocities in the PT range of the 
lower mantle.

This article is organized as follow: we first discuss the equations used for numerically 
determining the elastic constant tensor within the SC-QHA. 
We then describe the procedure for correcting it to first order for deviatoric thermal stresses. 
We then evaluate these corrections to the previously reported elastic constant tensors of 
PV\cite{RenataPRL} and PPV.\cite{Tsuchiya}

\section{Elasticity within and beyond the statically constrained (SC) QHA}

The present procedure builds on a related procedure to correct structural parameters and 
equations of state of non-isotropic solids at high PTs.\cite{Carrier}
The method introduced in Ref. \onlinecite{Carrier} can correct the SC crystal structure at $V(P,T)$ 
to infinite order as long as the SC elastic constant tensor is simultaneously corrected. 
However, this is a very demanding computational procedure and, fortunately, unnecessary. 
A first order correction to the crystal structure using SC elastic constant, appears to be 
sufficient. This conclusion was reached after examining the crystal structure of one of the most 
studied materials at high PT:  MgSiO$_3$-perovskite.\cite{ExpPV}
This type of experimental data is quite limited and results on other materials 
with similarly complex crystal structures would be helpful strengthen this conclusion.

According to the (SC) QHA the Helmholtz free energy is given by: 
\begin{eqnarray}
{F(V,T)} = \left[{E(V)} + \sum_{\textbf{q}j}\frac{\hbar\omega_{\textbf{q}j}({V})}{2} \right] 
+ k_B{T}\sum_{\textbf{q}j}\ln\left(1-e^{-\hbar\omega_{\textbf{q}j}({V})/k_B{T}}\right), \label{FofVT}
\end{eqnarray}
where $k_B$ and $\hbar$ are respectively Boltzmann's and Planck's constants.
The first term, $E(V)$, is the volume dependent static energy obtained after full structural 
relaxation under isotropic pressure, and $\omega(V)$ is the corresponding phonon spectrum.
Both phonon spectrum and static energy are here determined using the DFT
within the local density approximation (LDA),\cite{KohnSham} but the methodology 
is general and applicable to any first principles method. 
Structural relaxations are performed using a variable cell shape (VCS) algorithm\cite{VCS} 
and phonon spectra are computed using the PWscf code\cite{quantumespresso} as described 
in Ref.\ \onlinecite{SISSA},
based the linear response theory.
The second term in Eq.\ (\ref{FofVT}) is the zero point energy, $F_{ZP}$, such that the sum of 
the terms in the bracket is the energy at $T$=0 K.
The last term in Eq.\ (\ref{FofVT}) is the thermal excitation energy, $F_{th}$
(see Ref.\ \onlinecite{Anderson} for details).

Pressure, $P$, is obtained from $F$ using the standard thermodynamics relation
\begin{eqnarray}
{P}  = -\left. \frac{\partial {F}}{\partial {V}}\right|_{{T}}. \label{PofF}
\end{eqnarray}
This procedure implicitly assumes that $P$ remains isotropic at all temperatures, but this 
is only true for static calculations, where structures were optimized at target pressures.
The two frequency dependent terms in Eq.\ (\ref{FofVT}), the zero-point energy and the thermal 
energy, contribute to $P$ but their strain gradients are intrinsically anisotropic.
This effect was recently quantified\cite{Carrier} by the computation of deviatoric thermal 
stresses, $\delta \sigma_k$, defined as the difference between the stress tensor and the 
nominal pressure (diagonal) tensor. In Voigt's notation:
\begin{equation}
\delta \sigma_k = \left.\frac{1}{V_0}\frac{\partial {G(P,T)}}{ \partial\epsilon_k}\right|_{ \scriptsize {P}, {T}} 
- H(3-k){P},
\hspace{1cm} \mbox{for}\hspace{3pt} k=1,\dots,6, \label{eq:sigma}
\end{equation}
where $H(n)$ is the Heaviside step function, equal to 0 for $(3-k)$ strictly negative and 
1 otherwise.
Deviatoric thermal stresses are caused by the vibrational (zero-point and thermal) energies and 
are shown to be important at high pressures and temperatures. The larger the temperature, 
the more visible these stresses are.

We have previously shown that these deviatoric stresses can be relaxed to first order if one 
knows the elastic constant tensor, $c_{ij}({P}, {T})$, calculated within the (SC) 
QHA.\cite{Carrier} 
The latter are obtained from the Gibbs free energy, $G$,
\begin{equation}
{G}(P,T) = F+{P}{V} \label{eq:G}
\end{equation}
by calculating the second derivative of $G$ with respect to the strains $\epsilon_i$ and 
$\epsilon_j$:\cite{RenataPRL, Nye}
\begin{equation}
c_{ij}(P,T) = \frac{1}{V_0} \left.\frac{\partial^2 G}{\partial \epsilon_i\partial 
\epsilon_j}\right|_{T}. \label{eq:cij}
\end{equation}
The adiabatic elastic constants, which are the relevant ones for interpretation of seismic data, 
are then computed using appropriate thermodynamics relations.\cite{RenataPRL, Musgrave}
Below all calculated elastic constants, bulk and shear moduli, and velocities are adiabatic.

Lattice parameters at high pressures and temperatures under hydrostatic conditions can then be 
corrected to first order by evaluating the strains, $\epsilon_k$, involved in the relaxation 
of the deviatoric thermal stresses given in Eq.\ (\ref{eq:sigma}):
\begin{equation}
\epsilon_k({P}, {T}) = \sum_{m=1}^6 c_{km}^{-1}({P}, {T})\delta\sigma_m. \label{eq:deltaepsilon}
\end{equation}
The Cartesian components of the relaxed lattice vectors are then:  
\begin{equation}
\textbf{h}^{\star} = \textbf{h}(\textbf{I} - \mbox{\boldmath$\epsilon$}), 
\label{latticeCorrection}
\end{equation}
where
$$
\textbf{h} = \left( \begin{array}{ccc} a_{x} & b_{x} & c_{x} \\ a_{y} & b_{y} & c_{y} \\ 
a_{z} & b_{z} & c_{z} \end{array} \right) \hspace{10pt} \mbox{and} \hspace{10pt}
 \mbox{\boldmath$\epsilon$} = \left( \begin{array}{ccc} 
\epsilon_{1} & \epsilon_{6}/2 & \epsilon_{5}/2 \\ 
\epsilon_{6}/2 & \epsilon_{2} & \epsilon_{4}/2 \\ 
\epsilon_{5}/2 & \epsilon_{4}/2 & \epsilon_{3} \end{array} \right)
$$
are respectively the matrices of lattice vectors ($\vec{a}, \vec{b}, \vec{c}$) and Cartesian 
strains (keeping up with Voigt's notation).
Notice that increase in symmetry or symmetry break (phase transformations) may be induced 
by deviatoric thermal stresses in the presence of soft phonon, i.e., $\textbf{h}$ and 
$\textbf{h}^{\star}$ do \emph{not} necessarily have to the same space group.

In Ref.\ \onlinecite{Carrier} we pointed that attainment of zero deviatoric thermal stresses 
within the QHA should involve a \emph{self-consistent} cycle with simultaneous recalculation of 
the elastic constant tensor under hydrostatic condition followed by new structural relaxation, 
and so on. 
However, such procedure is extremely computationally intensive given the need to recompute
vibrational density of states on a PT grid every step of the cycle. 
We show next how to obtain the elastic constant tensor corrected to first order with knowledge 
of (\ref{eq:deltaepsilon}) only. 

The components of the elastic constant tensor expanded in a Taylor series of strains (in 
Voigt's notation) defined by Eq.\ (\ref{eq:deltaepsilon}) are:
\begin{eqnarray}
\lefteqn{c_{ij}(P, T,  \mbox{\boldmath$\epsilon$})  =  c_{ij}(P, T, \textbf{0}) +} \nonumber \\ & & 
 + \sum_{k=1}^6 \left.\frac{\partial c_{ij}}{\partial \epsilon_k}\right|_{P, T}  \epsilon_k + 
 \sum_{k=1}^6 \sum_{l=1}^6 \left.\frac{\partial^2 c_{ij}}{\partial \epsilon_k\partial 
\epsilon_l}\right|_{P, T}  \epsilon_k  \epsilon_l + \cdots \nonumber \\
\end{eqnarray}
Neglecting second and higher order terms one has:
\begin{eqnarray}
c_{ij}(P, T, \mbox{\boldmath$\epsilon$})  = & \simeq & c_{ij}(P, T) + 
\sum_{k=1}^6 \sum_{m=1}^6\left.\frac{\partial c_{ij}}{\partial 
P}\right|_{P, T}\left.\frac{\partial P}{\partial \sigma_m} \right|_{P, T} 
\left.\frac{\partial \sigma_m}{\partial \epsilon_k} \right|_{P, T} \epsilon_k, \nonumber \\ 
& = & c_{ij}(P, T) + \left.\frac{\partial c_{ij}}{\partial P}\right|_{P, T} 
\sum_{m=1}^3\left.\frac{\partial P}{\partial \sigma_m} \right|_{P, T} \delta \sigma_m.
\label{eq:Taylor}
\end{eqnarray}
In the last step above we assumed that pressure is unaffected by shear stresses, i.e., 
$\displaystyle \left. \frac{\partial P}{\partial \sigma_m}\right|_{P,T} = 0$ for $m=$ 4, 5, and 6, 
thus reducing the index summation from 6 to 3.
The stress derivatives of $P$ in Eq.\ (\ref{eq:Taylor}) are determined using the definition 
of the pressure as the trace of the stress tensor, ${\displaystyle P \equiv \frac{1}{3}\sum_{m=1}^3 
\sigma_m}$. 
Taking the derivative of the pressure as function of each stress leads to $\displaystyle 
\left. \frac{\partial P}{\partial \sigma_m}\right|_{P,T} = \frac{1}{3}$, for $m=$1, 2, and 3.
Therefore the first order corrected elastic constants at the strains  given by 
Eq.\ (\ref{eq:deltaepsilon}) is reduced to: 
\begin{equation}
c_{ij}(P, T, \epsilon_1, \epsilon_2, \epsilon_3)  =
c_{ij}(P, T) + \frac{1}{3}\left.\frac{\partial c_{ij}}{\partial P}\right|_{P, T}\delta 
\sigma_{\Sigma},
\label{equationCorrection}
\end{equation}
where $\displaystyle \delta \sigma_{\Sigma} = \sum_{k=1}^3 \delta \sigma_k$.
This correction requires only knowledge of the pressure derivatives of $c_{ij}$'s which are 
known from the statically constrained QHA calculation, and the deviatoric thermal stresses 
given by Eq.\ (\ref{eq:sigma}).
It gives to first order the elastic constants corrected for deviatoric stresses without having 
to explicitly calculate Gibbs free energy at the relaxed lattice parameters. 

As a final remark, we point that Eq.\ (\ref{equationCorrection}) could also be used and tested 
on experimental data as a mean for correcting any type of deviatoric stresses, as long as 
the stress deviations remain small compared to the hydrostatic pressure (in a limit for the 
Taylor expansion to be valid).
The correction only requires knowledge of (i) the three components $\delta \sigma_k$,  
$k=1,2,3$, and at the same time (ii) the pressure variation of the elastic constants at specified $P$ and 
$T$:  $\left.\frac{\partial c_{ij}}{\partial P}\right|_{P, T}$.
Principal strain deviations, $\epsilon_{\perp}$ and $\epsilon_{||}$, are measurable quantities, 
for instance, using diffraction ring measurements\cite{Bassett}
and their corresponding stresses are therefore also available from experiments. 
Pressure variation of the elastic constants\cite{Daniels} are measurable quantities\cite{Singh} 
that require only few additional runs for estimating experimentally the pressure derivative 
of $c_{ij}$  at given $PT$'s.
Eventual experimental setting that combines \emph{simultaneously} measurements of (i) and (ii) above
can be used to measure the correction to the elastic constants  due to
deviatoric stresses in DAC apparatus, after applying Eq.\ (\ref{equationCorrection}).

\section{Elastic constants of PV and PPV}

We present in this section new results on the deviatoric thermal stresses of PPV and the 
correction to the elastic constants obtained using the (SC) QHA.\cite{Tsuchiya}
Since deviatoric thermal stresses of PV were recently published,\cite{Carrier} 
we also give here the 
corresponding correction to the elastic constants of PV.

\begin{figure}[tb]
   \centering
   \includegraphics[width=3.0in]{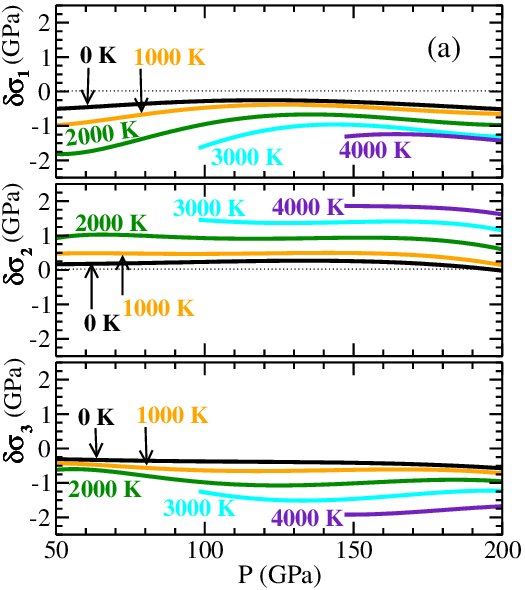} 
   \includegraphics[width=3.13in]{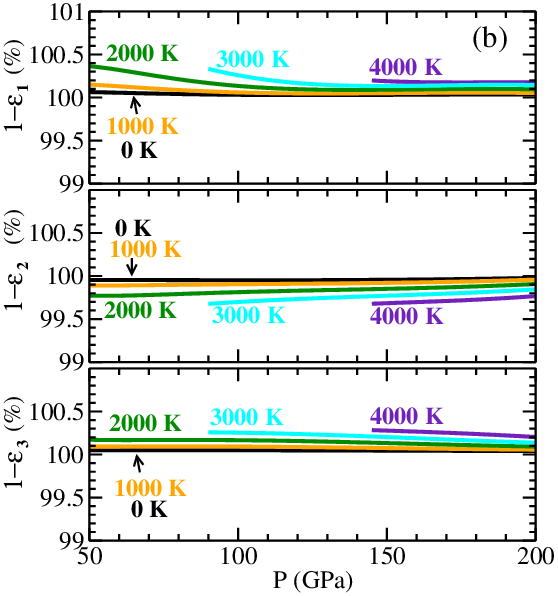} 
     \caption{(Color online) (a) Deviatoric thermal stresses in PPV; (b) percentage lattice constant corrections in PPV. 
$\delta \sigma_1$ and $\delta\sigma_2$ 
have opposite signs and similar magnitude, similarly to the case of  PV.\cite{Carrier} However, $\delta\sigma_3$ in PPV is considerably larger than in PV.}
   \label{Fig1:deltaSigPPV}
\end{figure}

The PT dependent elastic constant tensors of PV and PPV determined using 
the (SC) QHA have been reported respectively in Ref.\ \onlinecite{RenataPRL} and   
Ref.\ \onlinecite{Tsuchiya}. These are the major phases of the Earth's lower 
mantle and their elastic properties are central information for the interpretation 
of the seismic properties of this inaccessible region in terms of temperature, 
composition, and mineralogy. 
PV and PPV are both orthorhombic crystals respectively with symmetry $Pbnm$ and $Cmcm$.
This difference of symmetry group implies in particular, as stated in Ref. \onlinecite{Tsuchiya}, 
that
``the $[100]_{\mbox{PPV}}$, $[010]_{\mbox{PPV}}$, and $[001]_{\mbox{PPV}}$ directions in 
the $Cmcm$ structure 
correspond to the $[1\bar{1}0]_{\mbox{PV}}$, $[110]_{\mbox{PV}}$, and $[001]_{\mbox{PV}}$ 
in the $Pbnm$ structure, respectively,''
corresponding to a rotation of 45$^{\circ}$ of the $\vec{a}$---$\vec{b}$ reciprocal lattices.
Lattice deformations and deviatoric thermal stresses between PV and PPV are thus comparable only 
through this transformation.
Figure \ref{Fig1:deltaSigPPV}(a) shows the deviatoric thermal stresses for PPV. 
Equivalent results for PV have recently been reported in Ref.\ \onlinecite{Carrier} along 
with the analysis of its crystalline structure at high PT.
The deviatoric stresses $\delta\sigma_1$ and $\delta\sigma_2$ in PPV have opposite sign but 
similar magnitudes to that of PV (see Ref.\ \onlinecite{Carrier}), 
except along the $\vec{c}$ crystalline axes.
As stated above, deviatoric thermal stresses for PV and PPV induce distinct deformations
along lattices $\vec{a}$ and $\vec{b}$.
The deviatoric thermal stresses in the $z$ direction of PPV is considerably larger than the 
corresponding one in PV leading to larger corrections in PPV than in PV, as shown below.
Figure \ref{Fig1:deltaSigPPV}(b) shows the percentage corrections to the lattice parameters 
of PPV, based on Eq.\ (\ref{eq:deltaepsilon}).
Interestingly, Fig.\ \ref{Fig1:deltaSigPPV} shows that zero-point energy (the black zero 
Kelvin line in that figure) also produces deviatoric stresses. With increasing temperature, 
these stresses are enhanced but their origin is the anisotropic nature of the phonon dispersions.

\begin{figure}[tb] 
   \centering
   \includegraphics[width=3.87in]{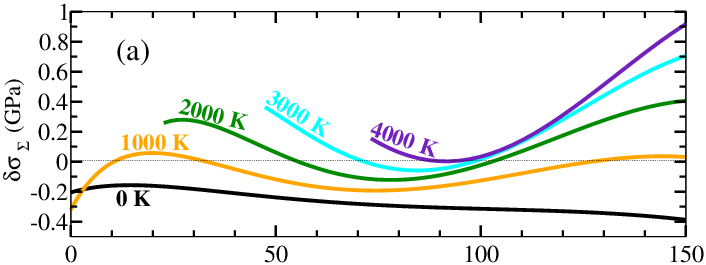}
    \mbox{}  \vspace{0.5cm} \mbox{}
   \includegraphics[width=3.8in]{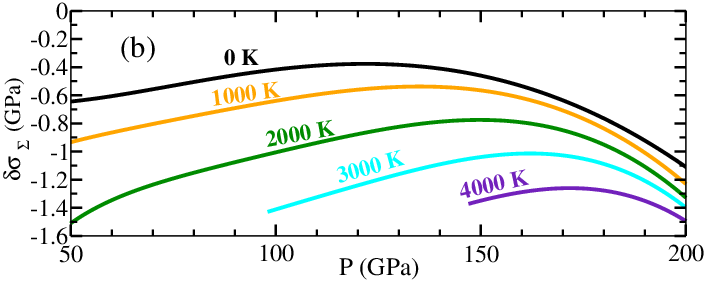} 
     \caption{(Color online) Sum of diagonal deviatoric stresses for (a) PV and (b) PPV, as defined in Eq.\ (\ref{equationCorrection}). 
      This sum is considerably larger in PPV because of the larger contribution from $\delta\sigma_3$ in PPV.
      Note that the pressure ranges 
     between PV and PPV differ, corresponding to their respective QHA regions of validity.\cite{Carrier} }
   \label{Fig2:SUMdeviatoricStress}
\end{figure}

Figure \ref{Fig2:SUMdeviatoricStress} shows the resulting summation of the three deviatoric 
thermal stresses $\delta \sigma_{\Sigma}$ [of Fig.\ \ref{Fig1:deltaSigPPV}(a)] for PPV 
(and see Ref.\ \onlinecite{Carrier} for PV's deviatoric thermal stresses).
It represents the first of the two ingredients necessary for the correction given by Eq.\ 
(\ref{equationCorrection}). 
Clearly, the correction for PPV is considerably larger than the one for PV. This is mostly due 
to $\delta\sigma_3$ that is larger in PPV than in PV (see above).
The correction for PPV is always negative, which has the effect of decreasing its elastic 
constants, while for PV, the correction can be negative (mostly at low temperature) or 
positive (mostly at high temperature). 
In principle, there are no reasons for having deviations of systematic nature and they should 
vary depending on the crystalline structure. 
One observation that remains true for all crystalline structures, however, is that positive 
deviations in one direction are to be compensated by a negative deviation in another 
direction, as observed in both PV and PPV. 

\begin{figure}[tb] 
   \centering
   \includegraphics[width=3.8in]{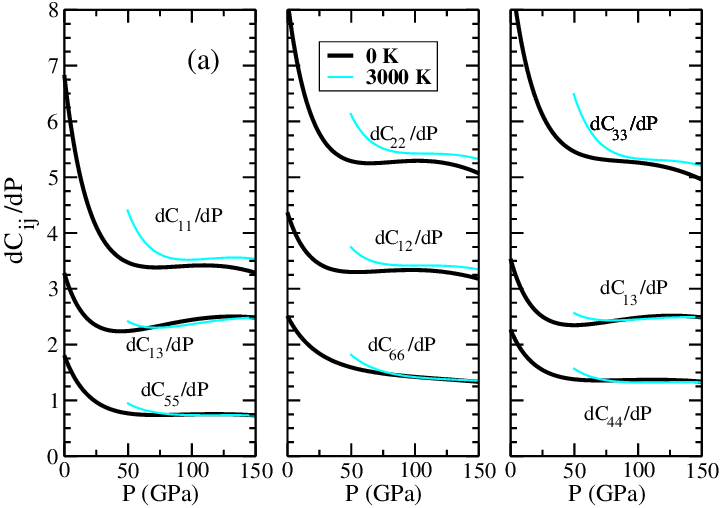} 
    \mbox{}  \vspace{1.2cm} \mbox{}
   \includegraphics[width=3.8in]{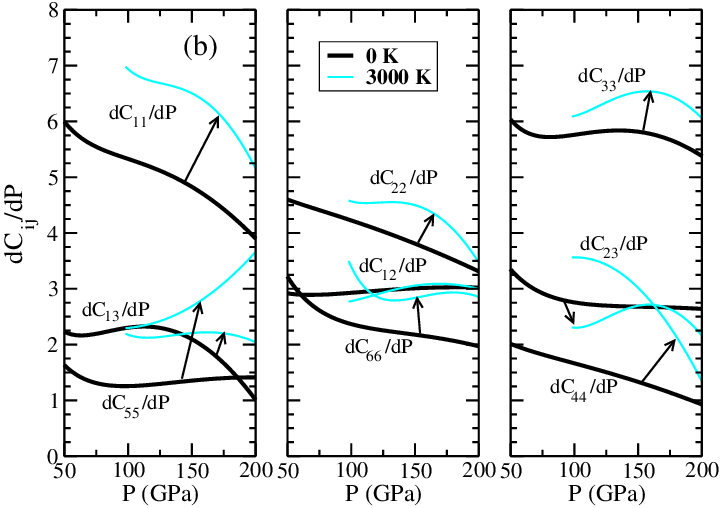} 
     \caption{(Color online) Derivatives of elastic constants as function of pressure of (a) PV and (b) PPV. 
     See also note in the caption of Fig.\ \ref{Fig2:SUMdeviatoricStress}.
}
   \label{Fig3:derivatives}
\end{figure}

Figure \ref{Fig3:derivatives} shows the pressure derivatives, $\partial c_{ij}/\partial P$, 
of all the elastic constants of PV and PPV, which is the second ingredient required for 
the correction according to Eq.\ (\ref{equationCorrection}). 
The figure shows the variations of $c_{ij}$ with pressure for only two temperatures, 
0 K and 3000 K, the latter being close to the temperature of the D$^{''}$ layer in the lower mantle,
where the PPV phase is important in the geophysical models.\cite{Lay} 

\begin{figure}[tb]
   \centering
   \includegraphics[width=3.8in]{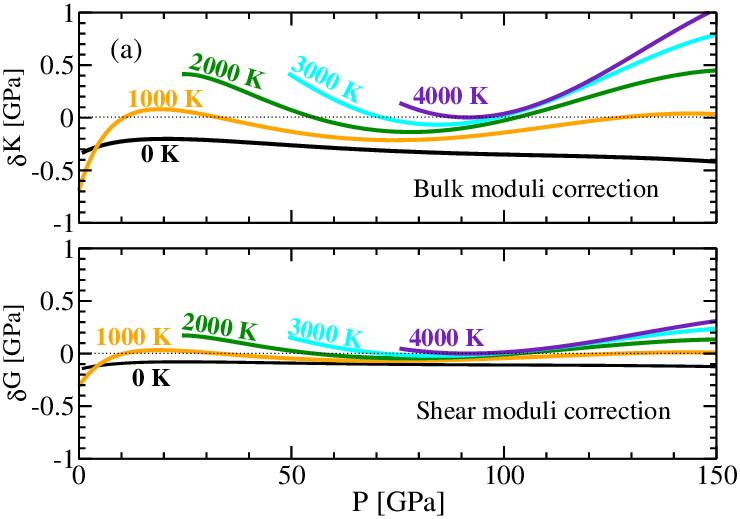} 
     \mbox{}  \vspace{1.2cm} \mbox{}
   \includegraphics[width=3.8in]{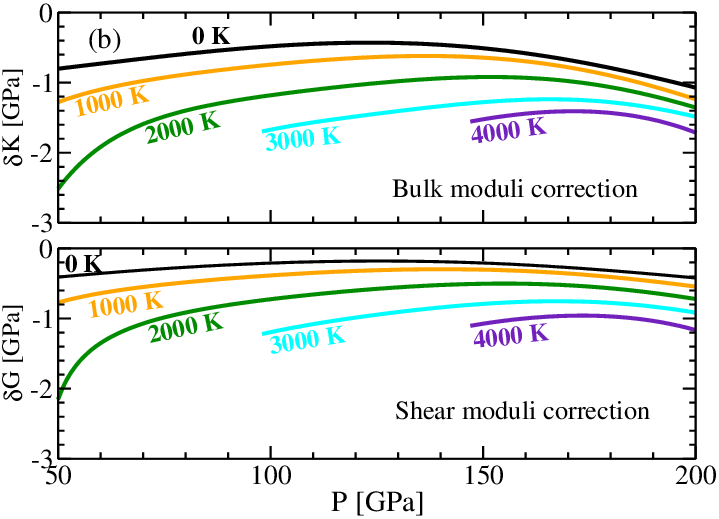} 
    \caption{(Color online)  Corrections to the bulk and shear moduli  for 
                   (a) PV\cite{RenataPRL} and (b) PPV.\cite{RenataPNAS}}
   \label{Fig4:KG}
\end{figure}

Figure \ref{Fig4:KG} shows the corrected bulk and shear moduli, after applying Hill's\cite{VRH}
(arithmetic) average to the elastic constants, at several temperatures. 
The corrections are largest at high pressure and high temperature in both PV and PPV. 
The nature of the correction is also structure-dependent.
Notice that the general aspect of the correction to the bulk moduli in Fig.\ \ref{Fig4:KG} is 
similar to $\delta \sigma_{\Sigma}$ displayed in Fig.\ \ref{Fig2:SUMdeviatoricStress}, 
indicating that the dominant term in the correction of Eq.\ (\ref{equationCorrection}) is the 
deviatoric thermal stress, and to a lesser extent the pressure derivatives of the elastic constants.
However, all corrections remain relatively small, meaning the (SC) QHA calculation does not 
suffer from significant deviatoric thermal stresses, although they can very well be 
corrected to any level of accuracy. 

Table \ref{Table1:data} summarizes the corrections to the (SC) QHA for the elastic constants 
at $T$ = 3000K for two pressures, $P$ = 100 GPa and $P$ = 120 GPa.
Corrections are given in parenthesis.
Bulk and shear moduli calculated using Voigt (uniform strain), Reuss (uniform stress), 
and Hill (arithmetic average between Voigt and Reuss) 
are shown.\cite{VRH}
The volume correction, $abc \times (1-\epsilon_1)(1-\epsilon_2)(1-\epsilon_3)$, as shown 
in Fig.\ \ref{Fig1:deltaSigPPV}(b), is reported as density, $\rho(P,T)$. 
Velocities are then evaluated from Voigt-Reuss-Hill moduli, since 
it provides a realistic estimation of the true moduli.
Notice that velocities are only slightly modified, because moduli are corrected along with the 
density; therefore, their ratio remains relatively unaltered.

\section{Conclusions}

In summary, we have introduced a scheme to correct high PT elastic constants obtained 
using the statically constrained quasiharmonic approximation for deviatoric thermal 
stresses that develop in calculations of anisotropic structures. This self-consistent scheme 
was used to compute to first order the elastic constants of the geophysically important 
MgSiO$_3$-perovskite and MgSiO$_3$-post-perovskite phases of the lower mantle. The corrections 
introduced by relaxation of these deviatoric stresses are quite small at relevant conditions 
of the lower mantle and previous (SC) QHA results remain essentially unchanged. However, this 
might not be the general case and the current scheme may be 
used to arbitrary order for computing high PT elastic constants to the desired level 
of accuracy.

\textbf{Acknowledgements} 

{This work was supported by NSF grants EAR-0230319, EAR-0635990, and ITR-0428774. 
We especially thank Shuxia Zhang from the Minnesota Supercomputing Institute for her assistance 
with optimizing the PWscf code performance on the BladeCenter Linux Cluster and on the SGI 
Altix XE 1300 Linux Cluster, and Yonggang Yu for helpful discussions relative to PWscf.
PC acknowledges partial support from a MSI research scholarship and JFJ from Brazilian 
agency CNPq.}

\newpage

\newpage\mbox{}\newpage

\newpage\mbox{}\newpage

\begin{table}[h]
\caption{\label{tab:table2}Elastic moduli of PV \cite{RenataPRL} and PPV 
\cite{RenataPNAS} with corrections given in parenthesis, as described by Eq.\ (\ref{equationCorrection}).
Pressure and elastic constants are in GPa, velocities in km/s, temperature in K, densities, $\rho$, in g/cm$^3$.
The corrections are significant for bulk and shear moduli. Velocities are only slightly changed by the correction. 
$V_P = \sqrt{(K^H + 4/3 G^H)/\rho}$, $V_S = \sqrt{G_H/\rho}$, $V_{\Phi} = \sqrt{K^H/\rho}$, and $\Phi = K^H/\rho$, where upper indeces $R$, $V$, and $H$ represent Reuss, Voigt and the Hill averages.\cite{VRH} Notice that $\Phi = V_P^2 - 4/3 V_S^2$. Velocities are calculated using the Hill averages. Decimal digits are presented to show the magnitude of the corrections. However, except for $\rho$, the accuracy of results should not include decimal digits. 
}
\begin{ruledtabular}
\begin{tabular}{crcrrcr}
                            & \multicolumn{3}{c}{3000 K, 100 GPa} & \multicolumn{3}{c}{3000 K, 120 GPa}  \\
                                \cline{2-4}                                 \cline{5-7}  
                            & PV \ \ \  \    & & PPV \ \ \  \     & PV \ \ \  \      & & PPV  \ \  \  \  \\
$c_{11}$                    &   774.8  (0.0) & &   933.4  (-3.3) &    844.4  (0.3) & &   1069.5  (-2.8) \\
$c_{22}$                    &   941.7  (0.0) & &   756.0  (-2.1) &   1049.9  (0.5) & &    846.8  (-1.9) \\
$c_{33}$                    &   928.5  (0.0) & &   949.0  (-2.9) &   1034.9  (0.5) & &   1072.5  (-2.6) \\
$c_{44}$                    &   287.2  (0.0) & &   215.8  (-1.7) &    313.6  (0.1) & &    286.6  (-1.4) \\
$c_{55}$                    &   251.0  (0.0) & &   164.4  (-1.1) &    265.6  (0.1) & &    211.1  (-1.0) \\
$c_{66}$                    &   248.4  (0.0) & &   253.3  (-1.6) &    276.4  (0.1) & &    314.9  (-1.2) \\
$c_{12}$                    &   452.7  (0.0) & &   376.7  (-1.3) &    520.4  (0.3) & &    433.4  (-1.2) \\
$c_{13}$                    &   373.9  (0.0) & &   370.6  (-1.0) &    421.3  (0.2) & &    413.3  (-0.9) \\
$c_{23}$                    &   406.5  (0.0) & &   434.3  (-1.1) &    455.7  (0.2) & &    481.1  (-1.0) \\ \cline{2-4}  \cline{5-7}  
$K^V$                       &   567.9  (0.0) & &   555.7  (-1.7) &    636.0  (0.3) & &    627.2  (-1.5) \\
$\textbf{K}^{\textbf{H}}$   &   565.2  (0.0) & &   553.3  (-1.7) &    632.2  (0.3) & &    624.3  (-1.5) \\
$K^R$                       &   562.4  (0.0) & &   550.8  (-1.7) &    628.4  (0.3) & &    621.5  (-1.5) \\ \cline{2-4}  \cline{5-7}  
${G}^{{V}}$                 &   251.4  (0.0) & &   223.8  (-1.2) &    273.3  (0.1) & &    273.3  (-1.0) \\
$\textbf{G}^{\textbf{H}}$   &   249.2  (0.0) & &   219.5  (-1.2) &    270.1  (0.1) & &    268.6  (-1.0) \\ 
$G^R$                       &   247.0  (0.0) & &   215.2  (-1.2) &    267.0  (0.1) & &    263.9  (-1.0) \\ \cline{2-4}  \cline{5-7}  
$\rho$                      &     5.04 (0.00) & &     5.11 (-0.01) &      5.22 (0.00) & &      5.29 (-0.01) \\
$\textbf{V}_{\textbf{P}}$   &    13.35 (0.00) & &    12.87 (-0.01) &     13.79 (0.01) & &     13.63 (-0.01) \\
$\textbf{V}_{\textbf{S}}$   &     7.03 (0.00) & &     6.56 (-0.01) &      7.20 (0.00) & &      7.13 (-0.01) \\
$\textbf{V}_{\Phi}$         &    10.59 (0.00) & &    10.41 (-0.01) &     11.01 (0.00) & &     10.86 ( 0.00) \\
$\Phi$                      &   112.20 (0.02) & &   108.33 (-0.10) &    121.21 (0.06) & &    118.02 (-0.11) \\
\end{tabular}
\end{ruledtabular}
\label{Table1:data}
\end{table}

\end{document}